# Accelerating Sparse Linear Solvers with an Optical Laser Processing Unit
# Invited Paper[1]

Dan Gluck

LightSolver Ltd., Tel Aviv, Israel, dani@lightsolver.com

Yotam Mimran

LightSolver Ltd., Tel Aviv, Israel, yotam@lightsolver.com

Andrey Karenskih

LightSolver Ltd., Tel Aviv, Israel, andreyk@lightsolver.com

Talya Vaknin

LightSolver Ltd., Tel Aviv, Israel, talya@lightsolver.com

Omri Wolf

LightSolver Ltd., Tel Aviv, Israel, omri@lightsolver.com

Ruti Ben-Shlomi

LightSolver Ltd., Tel Aviv, Israel, ruti@lightsolver.com

Johannes Gebert

Numerical Methods and Libraries, High-Performance Computing Center Stuttgart (HLRS), University of Stuttgart, Nobelstraße 19, 70569, Stuttgart, Germany, gebert@hlrs.de

**ABSTRACT**

Solving large, sparse linear systems is a fundamental workload in scientific computing and engineering simulations, often dominating runtime and energy consumption in high-performance computing (HPC) applications. In this work, we explore an alternative computing paradigm based on analog optical processing, implemented through the Laser Processing Unit (LPU). The LPU encodes linear systems into the dynamics of coupled lasers within an optical cavity, where the steady-state phases of the optical fields correspond to the solution of $Ax = b$.

We present a mapping of general linear systems, both dense and sparse, onto the LPU architecture and evaluate its performance using representative matrices from the SuiteSparse collection. Using an LPU emulator, we benchmark convergence behavior and time-to-solution for sparse, multi-banded matrices against established Krylov subspace methods (CG, GMRES, BiCGSTAB, and others) executed on a modern GPU platform. Our results demonstrate that the LPU will achieve significantly lower time-to-solution for selected problem classes, highlighting the potential of optical analog computing for accelerating iterative linear solvers.

These findings suggest that optical processors such as the LPU will be able to serve as accelerators for linear systems, in particular structured and/or repeatedly solved, offering advantages in latency, parallelism, and energy efficiency. We discuss current limitations, including scaling constraints and precision considerations, and outline directions toward hybrid optical–digital computing systems.

# 1 INTRODUCTION

Solving systems of linear equations is a central computational task across a wide range of scientific and engineering domains, including computational fluid dynamics, electromagnetics, structural analysis, and materials science. These applications frequently require the solution of large, sparse linear systems of the form $Ax = b$, where $A \in \mathbb{R}^{n \times n}$ is typically sparse and structured, $b \in \mathbb{R}^n$ is a known vector, and $x \in \mathbb{R}^n$ is the unknown solution. For many workloads, particularly those arising from discretized partial differential equations, solving such systems dominates the overall computational cost.

On conventional digital hardware, sparse linear systems are commonly solved using iterative Krylov subspace methods such as Conjugate Gradient (CG), GMRES, and BiCGSTAB. While these methods are highly optimized, their performance is often limited by memory bandwidth and data movement rather than raw arithmetic throughput. This "memory wall" is especially pronounced in sparse computations, where irregular memory access patterns reduce cache efficiency and limit scalability on modern architectures.

Analog and non-von-Neumann computing paradigms have recently re-emerged as promising alternatives for specific computational kernels [1,2,3]. In particular, optical computing offers unique advantages due to its inherent parallelism, high bandwidth, and ability to naturally perform linear transformations through interference and propagation [4,5]. These properties make optical systems especially attractive for accelerating linear algebra operations, where matrix–vector products can be embedded directly into physical processes. These advantages are particularly compelling for structured problems (such as multi-banded systems) that arise repeatedly in real-time pipelines, where time-to-solution, throughput, and power are often more important than floating-point exactness.

In this work, we investigate the Laser Processing Unit (LPU), an analog optical computing device that solves linear systems by exploiting the dynamics of coupled lasers in a degenerate cavity [6], i.e., an optical cavity that supports many independent spatial modes, enabling parallel evolution of multiple optical states. We present a novel mapping of the system of linear equations onto the lasers within the cavity. The key idea is to encode the coefficients of the linear system into the coupling between optical modes, such that the system evolves toward a steady state corresponding to the solution. Unlike digital iterative solvers or computations on photonic-chips [7], which explicitly perform discrete updates, the LPU performs computation through continuous-time physical evolution, effectively implementing an analog iterative process similar to Richardson iteration.

The motivation for this approach is twofold. First, by embedding the linear operator into the physical system, the LPU minimizes data movement, addressing a key bottleneck in modern HPC systems. Second, optical systems provide massive spatial parallelism and ultrafast signal propagation, enabling low-latency computation with potentially favorable energy efficiency, especially when the same physical transform is reused across multiple solves.

To evaluate the practical potential of this approach, we map sparse linear systems from the SuiteSparse collection [8,9] onto the LPU framework and benchmark its performance using an emulator. We compare convergence time and solution accuracy against a range of Krylov solvers implemented in the Ginkgo library [10] and executed on a high-end GPU. The selected benchmark problems span different domains and matrix characteristics, providing insight into the applicability of the LPU across diverse workloads.

The contributions of this paper are as follows:
- We present a formulation for solving general linear systems on the LPU using phase-encoded optical states.
- We demonstrate the mapping between the physical dynamics of coupled lasers and iterative linear solvers.
- We provide a comparative benchmark against state-of-the-art GPU-based Krylov methods using SuiteSparse multi-banded matrices.

Overall, this work positions the LPU as a promising accelerator for linear algebra workloads, supporting both dense and sparse, multi-banded matrices, in particular where latency, throughput, and energy efficiency are critical.

## 2 LIGHTSOLVER'S LPU

The LPU is based on a degenerate cavity ring laser, using 4f telescopes for its implementation. As shown in figure 1(a), all planes in the setup are relayed and imaged onto themselves with unit magnification and minimal distortion. By placing a thin gain medium at a specific plane, each point in this plane acts independently, allowing us to write field rate equations for the degenerate cavity (for more details see [4]).

$$\tau \frac{dE_i(t)}{dt} = e^{G_i(t)-\alpha} K_{ii} E_i(t) + \sum_{j \neq i} e^{G_j(t)-\alpha} K_{ij} E_j(t) - e^{G_i(t)-\alpha} E_i(t) \quad (1)$$

$$\tau_G \frac{dG_i(t)}{dt} = P - 2G_i(t)(1+|E_i|^2) \quad (2)$$

$E_i(t)$ represents the electric field in the cavity at a spatial point denoted by $i$. $K_{ii}$ is the self- coupling coefficient and $K_{ij}$ are the cross-couplings between the fields. Note that these couplings could be complex. $\alpha$ is the round-trip loss (from mirror imperfections, output coupling, etc.), $G_i$ is the gain of field $E_i$, acting as a non-linear amplifier to compensate for the loss and ensure uniform steady-state amplitude, and $\tau$ is the round-trip time. $\tau_G$ and $P$ are parameters related to the gain dynamics.

The steady state of these equations, if exists, will be the state for which the lasers will realize minimal loss, thus, it depends on the coupling coefficients $K_{ij}$ and $K_{ii}$.

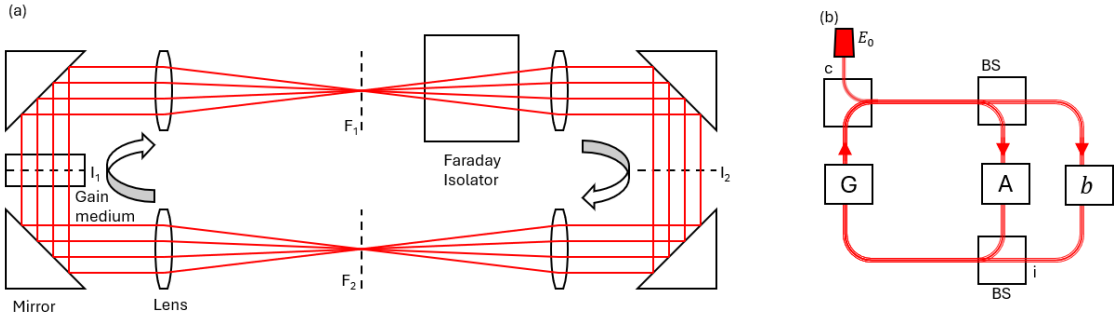

Figure 1: (a) Degenerate ring cavity laser. A 2D plane is imaged onto itself after traversing the cavity, while the gain medium is positioned in this image plane. Without additional components, each diffraction-limited spot will have a physical state independent of other spots. The Faraday isolator ensures unidirectional beam propagation. Each image (or Fourier) plane can be replaced by an additional 4f telescope to add functionality. (b) Schematic of an $Ax = b$ solver. An external field is introduced into the cavity lasers via couplings $c_i$. Here, $A$, $b$ and $c$ in the diagram are optical couplers implementing the system of linear equations, G is the gain medium, $E_0$ are the reference lasers and BS are beam splitters, also used to split and combine lasers. "i" stands for a 90-degree phase addition to the laser, which can be implemented in the optical coupler b.

## 3 SOLVING LINEAR EQUATIONS USING THE LPU

In this section we show how the LPU can solve a set of linear equations $Ax = b$; $A \in \mathbb{R}^{n \times n}$, $x, b \in \mathbb{R}^n$. We map the system of equations to the phases of $n + 1$ lasers, where one laser is a reference laser and the value $x_i$ is identified with the relative phase of the $i^{th}$ laser to the reference laser. A rescaling of the values is also required to obey the physical limitations of the LPU. We denote the problem lasers as $E_i$, for $1 \leq i \leq n$. The reference laser is denoted $E_0$.

We assume that the gain and loss are such that all fields have the same amplitude at steady state. We thus write:

$$E_j(t) = De^{i\varphi_j} \quad (3)$$

with constant amplitude $D$. The solution vector $x_i$ will be encoded on the phase $\varphi_i$ as will be seen below. Let us denote by $g(|E_i|)$ the gain and loss function for field $E_i$:

$$g(|E_i|) = e^{G_i(t) - \alpha} \quad (4)$$

We write the following equation for $E_i$ where coefficients $a_{ij}$, $b_i$ are elements from $A$ and from $b$ respectively. They are incorporated through coupling between lasers as follows:

$$\frac{dE_i}{dt} = \sum_j a_{ij} g(|E_j|) E_j - i b_i g(|E_i|) E_i + c_i g(|E_0|) E_0 \quad (5)$$

Where it includes contributions from the other fields, according to the matrix $A$, an imaginary contribution to itself, according to $b_i$ and contribution from the reference lasers according to

$$c_i \equiv -\sum_j a_{ij} \quad (6)$$

(See figure 1(b)). Thus, we have:

$$i D e^{i\varphi_i} \dot{\varphi}_i = g(|D|) \left[ \sum_j a_{ij} D e^{i\varphi_j} - i b_i D e^{i\varphi_i} + c_i D e^{i\varphi_0} \right] \quad (7)$$

Equivalently, the equation for the phase can be written explicitly as

$$i \dot{\varphi}_i = g(|D|) \left[ \sum_j a_{ij} e^{i(\varphi_j - \varphi_i)} - i b_i + c_i e^{i(\varphi_0 - \varphi_i)} \right] \quad (8)$$

Considering the imaginary and real parts of the equation separately, at its steady state, yields the following equations:

$$0 = \sum_j a_{ij} \sin(\varphi_j - \varphi_i) - b_i + c_i \sin(\varphi_0 - \varphi_i) \quad (9)$$

$$0 = \sum_j a_{ij} \cos(\varphi_j - \varphi_i) + c_i \cos(\varphi_0 - \varphi_i) \quad (10)$$

In the small angle approximation, i.e. $\varphi_j \approx \varphi_i$ and $\varphi_i \approx \varphi_0$ (the latter condition is only required if $\sum_j a_{ij} \neq 0$) the equations can be combined using (6) to yield the following result:

$$0 \approx \sum_j a_{ij} (\varphi_j - \varphi_0) - b_i \quad (11)$$

By identifying $x_i \equiv \varphi_i - \varphi_0$, we obtain the vector $x$ that solves for $Ax = b$. The small angle approximation can be ensured to hold by rescaling $b$ and $A$.

Note, that $A$ can be any general square matrix, either dense or sparse, including complex matrices, although as in any iterative method convergence is not always guaranteed. When $A$ is a sparse, multi-banded matrix, the implementation of the optical couplers in the LPU is especially efficient, so that a single LPU can tackle relatively large such matrices, examples of which are mentioned next.

## 4 METHODS

### 4.1 Problem selection

The problems used in this study are selected from the SuiteSparse library [8,9]. We specifically chose multi-banded problems that could fit on a single GPU and are projected to fit on a single LPU device. Three representatives were chosen, giving a variety of matrix properties and physical origin:

- **EPB3** - Epb3 is a non-symmetric, non-positive definite thermal problem, with $n = 84{,}617$ and 463,625 nonzero entries. It originates from a plate-fin heat exchanger model. This matrix is indicative of a class of non-symmetric problems arising in engineering applications.
- **Xenon2** - Xenon2 is a non-symmetric, non-positive definite materials problem, with $n = 157{,}464$ and 3,866,688 nonzero entries. It originates from a materials or nuclear-related simulation. This matrix represents a broader class of challenging non-symmetric systems encountered in multi-physics applications.

- **BenElechi1** - BenElechi1 is a symmetric positive definite 2D/3D problem, with $n = 245{,}874$ and 13,150,496 nonzero entries This matrix is indicative of a wide range of symmetric sparse systems arising in different applications such as structural mechanics.

The matrices were provided in the CSR (Compressed Sparse Row) format, and the vector b was generated randomly for each instance.

### 4.2 Benchmarking solvers

To evaluate the anticipated performance of the LPU, we compared its results against various solvers using the Ginkgo framework [10]. The solvers included in the benchmark were the following Krylov methods: CG, GMRES, Bicgstab, minres, idr, fcg, cgs, and bicg. These were run on NVIDIA GeForce RTX 3090 GPU.

Each solver was run ten times, with the initial state for all runs set to zeros, and the residual norm was calculated according to $\|Ax - b\|/\|b\|$. The tolerance is set to $1e - 5$, which is the tolerance reached by the LPU emulator.

For the LPU, convergence time was measured as the number of roundtrips multiplied by 20 nanoseconds, which represents the expected duration for a roundtrip within the cavity.

## 5 RESULTS

The bar charts below (figures 2-4) show the time to solve the different problems and different solvers. For each solver and each problem, the convergence time reported corresponds to the median of the actual solving durations ("apply time"). Error bars indicate the range between the 25th and 75th percentiles.

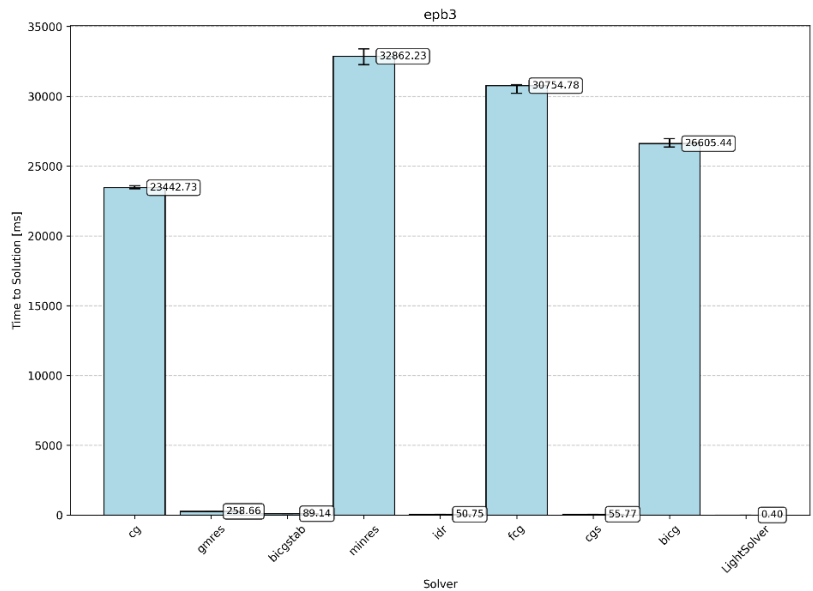

Figure 2: Solve runtimes for epb3

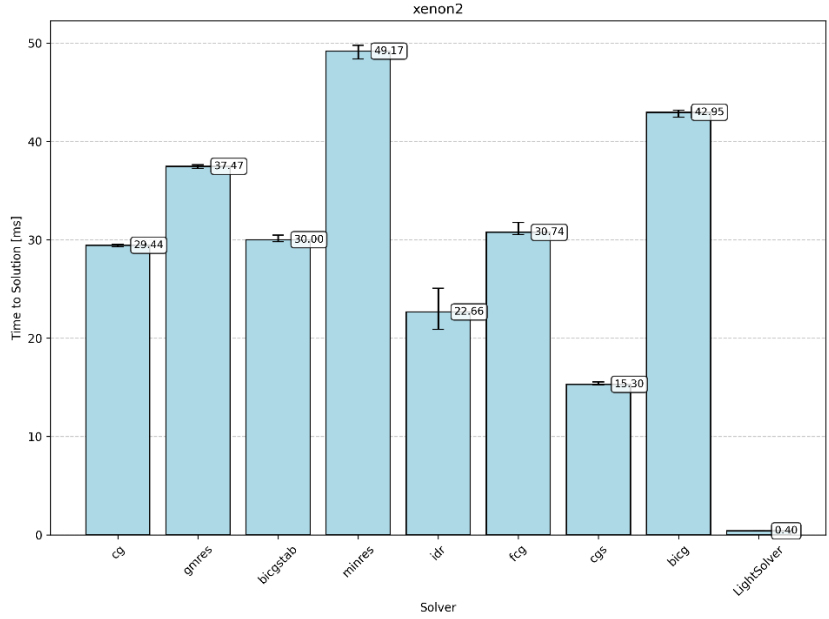

Figure 3: Solve runtimes for Xenon2

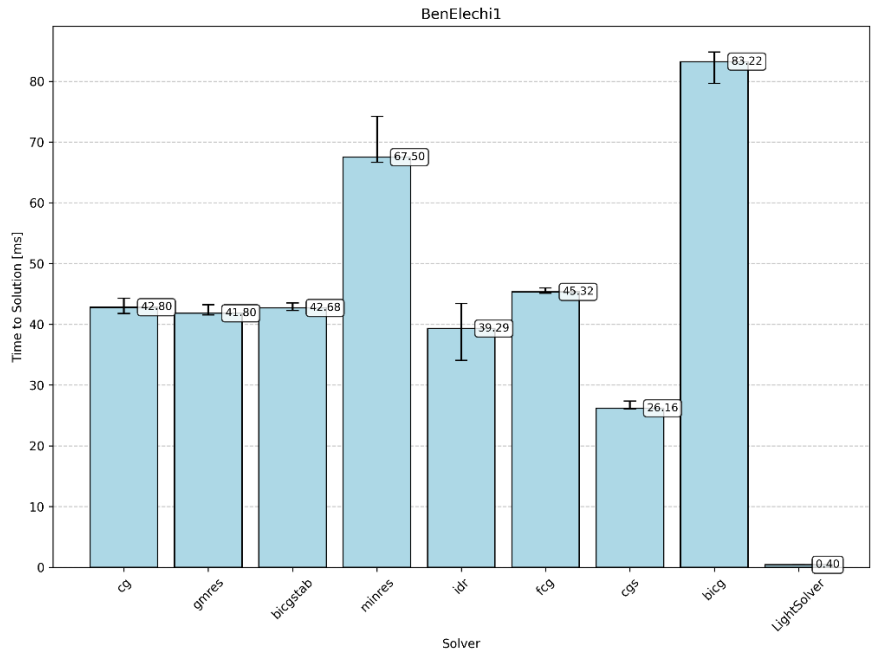

Figure 4: Solve runtimes for BenElechi1

## 6 SUMMARY AND OUTLOOK

In this work, we presented an optical analog approach for solving general linear systems using the Laser Processing Unit (LPU). By embedding the linear system directly into the coupling dynamics of a degenerate cavity laser, the LPU computes solutions through its natural convergence to a steady state. This paradigm differs fundamentally from conventional digital solvers, replacing discrete iterations with continuous physical evolution.

Using representative multi-banded matrices from the SuiteSparse collection, we evaluated the expected performance of the LPU via an emulator. We compared it against established Krylov subspace methods executed on a GPU platform. The results (see Figures 2–4 in the paper) indicate that the LPU achieves competitive convergence times and, for certain problem classes, will significantly outperform digital solvers in terms of latency. These advantages stem primarily from the intrinsic parallelism of the optical system and the elimination of repeated data movement or opto-electronic conversion within the computation loop.

A promising next step is the development of hybrid optical–digital workflows in which the LPU serves as a specialized accelerator rather than a standalone replacement for conventional solvers. In such systems, the digital processor would handle preprocessing, scaling, matrix normalization, residual evaluation, and fallback handling, while the optical hardware would execute the computationally dominant process of computing the solution to the set of linear equations. This division is especially attractive for iterative methods where intermediate solutions are valuable. For example, using the LPU as a fast initializer or an inner solver within a larger Krylov-based scheme.

The same approach can be extended to large-scale problems: although many real-world applications involve linear systems with billions of unknowns that exceed the capacity of a single LPU (or GPU), the LPU can still contribute within hierarchical or iterative workflows by approximately solving block-diagonal or domain-decomposed subproblems and thereby acting as a preconditioner inside a Krylov solver. In this setting, the LPU accelerates computationally intensive components while overall solution progress is orchestrated by conventional digital methods; exploring such hybrid large-scale strategies is beyond the scope of the present work.

At the same time, several challenges remain. The current approach relies on approximations (e.g., small-angle assumptions) and requires normalization of the problem to match physical constraints. Precision is inherently limited by the analog nature of the system, and scaling to larger or less structured matrices introduces additional complexity in encoding and control. Furthermore, the present evaluation is based on an emulator, and experimental validation on physical hardware will be essential to fully characterize performance, noise sensitivity, and robustness.

Looking forward, several directions appear promising:
- **Hybrid optical–digital workflows:** using the LPU as a fast preconditioner or accelerator within conventional iterative methods.
- **Problem specialization:** the LPU, by design, handles both sparse and dense matrices on the same hardware. Future research can evaluate sparse and dense computations without altering the hardware, thereby demonstrating their viability for either HPC or AI-specific mathematical objects.
- **Hardware scaling:** increasing the number of optical modes and improving control over coupling coefficients.
- **Error mitigation:** incorporating feedback or correction mechanisms to improve solution accuracy.

In conclusion, the LPU demonstrates the potential of optical analog computing as an alternative approach to traditional HPC architectures. It offers a compelling path forward for accelerating specific classes of linear algebra workloads that are central to modern scientific computing.